\documentclass[trackchanges,twocolumn]{aastex701}
\usepackage{amsmath}

\begin{document}

\title{Electron acceleration by turbulent reconnection in solar flares}

\author[orcid=0009-0007-8418-1986]{Zining Ren}
\affiliation{School of Astronomy and Space Science, Nanjing University, Nanjing 210023, People's Republic of China}
\affiliation{Key Laboratory for Modern Astronomy and Astrophysics (Nanjing University), Ministry of Education, Nanjing 210023, People's Republic of China}
\email{rzn@smail.nju.edu.cn}  

\author[orcid=0000-0003-2837-7136]{Xin Cheng} 
\affiliation{School of Astronomy and Space Science, Nanjing University, Nanjing 210023, People's Republic of China}
\affiliation{Key Laboratory for Modern Astronomy and Astrophysics (Nanjing University), Ministry of Education, Nanjing 210023, People's Republic of China}
\email[show]{xincheng@nju.edu.cn}

\author[orcid=0000-0001-9863-5917]{Yulei Wang}
\affiliation{Institute of Science and Technology for Deep Space Exploration, Suzhou Campus, Nanjing University, Suzhou, 215163, People's Republic of China}
\affiliation{State Key Laboratory of Lunar and Planetary Sciences, Macau University of Science and Technology, Macau, People's Republic of China}
\email[show]{wyulei@nju.edu.cn}

\author[orcid=0000-0002-4978-4972]{Mingde Ding}
\affiliation{School of Astronomy and Space Science, Nanjing University, Nanjing 210023, People's Republic of China}
\affiliation{Key Laboratory for Modern Astronomy and Astrophysics (Nanjing University), Ministry of Education, Nanjing 210023, People's Republic of China}
\email{dmd@nju.edu.cn}

\begin{abstract}
Solar flares can release magnetic energy explosively in the corona and produce high-energy particles on short timescales.
However, how and where these particles are accelerated remains an open question.
Here, we investigate the acceleration and transport of electrons during self-developed three-dimensional turbulent reconnection of solar flares by solving Parker’s transport equation in the framework of high-resolution MHD simulations.
We find that thermal electrons at both the flare current sheet and loop top are rapidly accelerated up to $\sim90\,\mathrm{keV}$, with energy spectra exhibiting a power-law feature. 
Although the population of accelerated electrons at the flare loop top is larger than that at the current sheet, their spectral indices are similar, close to the values usually observed. 
More importantly, the acceleration is achieved by turbulence-driven compression structures of various scales rather than the supposed termination shock, particularly at the flare loop top. 
A portion of compression structures even forms shocks. 
These results highlight the critical role of turbulent reconnection in accelerating electrons, thereby shedding new light on the acceleration and transport of particles in other high-energy phenomena.
\end{abstract}

\keywords{\uat{Solar flares}{1496} --- \uat{Solar magnetic reconnection}{1504} --- \uat{Solar energetic particles}{1491} --- \uat{Magnetohydrodynamical simulations}{1966} --- \uat{Solar physics}{1476}}

\section{Introduction} 
Particle acceleration is a fundamental physical process occurring in explosions of various celestial scales, ranging from magnetic storms around the Earth to solar and stellar flares, and even to distant supernova remnants and fast radio and gamma-ray bursts \citep{2007zank,2011Zharkova,1987Blandford,2016book}.
As the most energetic phenomena in the solar system, solar flares can release up to $\sim10^{32}$ erg of energy within minutes and accelerate vast numbers of particles, such as electrons to tens of keV and protons from MeV to GeV \citep{Fletcher,Emslie2012,2017Aschwanden}, posing significant risks to the safety of spacecraft and human activities in outer space.

Observationally, nonthermal emissions from accelerated electrons are mainly detected at the flare loop-top and footpoint regions, providing important clues to where and how particle acceleration may operate in solar flares.
Hard X-ray (HXR) sources have been frequently observed at the loop top since the pioneering Yohkoh discovery by \citet{1994Masuda}, providing strong evidence that energetic electrons are accelerated and trapped in this region \citep{2015oka,2020Yu}.
This loop-top emission is generally interpreted as the combined result of a magnetic bottle structure configuration \citep{2021chen} and a termination shock formed where reconnection outflows collide with the underlying flare loops \citep{2015Scichen}.
These energetic particles can subsequently stream downward along the flare loops toward the footpoints, where they lose energy through Coulomb collisions, producing HXR footpoint sources \citep{1981hoyng} and driving chromospheric evaporation that fills the flare loops with hot plasma, observed as soft X-ray (SXR) flare loops \citep{2008krucker}.

A variety of mechanisms have been proposed to explain the acceleration of particles during solar flares, such as the direct current (DC) electric field acceleration \citep{1996Litvinenko,1985Holman}, the plasmoid acceleration \citep{2010oka,2013Drake}, the stochastic acceleration \citep{2012Petrosian,2023ruan}, and the diffusive shock acceleration (DSA) \citep{1998Tsuneta,2012guo}. 
In particular, high-cadence radio imaging suggests that the loop-top termination shock plays a crucial role in particle acceleration \citep{2015Scichen,2021chen}. 
Nevertheless, how and where the energetic particles are accelerated during solar flares remains debated.

The primary challenge of this problem lies in the multi-scale nature of the acceleration process, spanning from kinetic scales on the order of meters to macroscopic scales exceeding hundreds of megameters. 
This scale disparity hinders comprehensive cross-scale observations and first-principles particle-in-cell (PIC) simulations \citep{2017li,2017Dahlin,2024zhang}.
To bridge this gap, hybrid approaches that couple large-scale magnetohydrodynamic (MHD) simulations with particle dynamics have been developed \citep{2015bai,2023MNRASsun,2025arXivhu,2021arnold,2025RAAli,2025ApJwu,2024Bacchini,2026Mora}.
Among these, particle transport models based on Parker’s transport equation (PTE) or focused transport equations \citep{parker1965,2014zank} provide an efficient framework to investigate particle acceleration and transport in global flare environments and have been used to reproduce observations successfully \citep{2019kong,2022kong,2024chen}.
The results seem to support the termination-shock acceleration scenario.

However, the termination-shock scenario does not capture how dynamically evolving turbulence, as recently resolved in flare current sheet (CS) and loop top \citep{cx2018,2018warren}, may distort the quasi-2D shock surface and alter the acceleration process.
High-resolution three-dimensional (3D) MHD simulations also showed that, due to strong turbulence, the termination shock formed during the early stage of the reconnection will be disrupted and replaced by ubiquitous fragmented shocks in 3D \citep{2022shen,wangyulei2023}. 
Whether such fragmented loop-top shocks can still accelerate electrons efficiently has not yet been assessed.
Additionally, the observed non-thermal emissions at the long-stretched CS cannot be explained adequately by the termination-shock model \citep{2018gary,2020NAchenbin,2022kou}, further motivating the search for new mechanisms.

In this work, we solve PTE within the framework of a high-resolution 3D MHD simulation of turbulent reconnection during solar flares. 
The important finding is that thermal electrons at both the flare CS and loop top can be rapidly accelerated to a power-law distribution by turbulence-driven compression structures (including fragmented shocks), differing from the termination shock previously insisted upon, particularly at the flare loop top.

\section{Numerical model} \label{sec:numerical model}

\subsection{MHD simulation of turbulent reconnection}
\begin{figure}
\centering
\resizebox{\hsize}{!}{\includegraphics{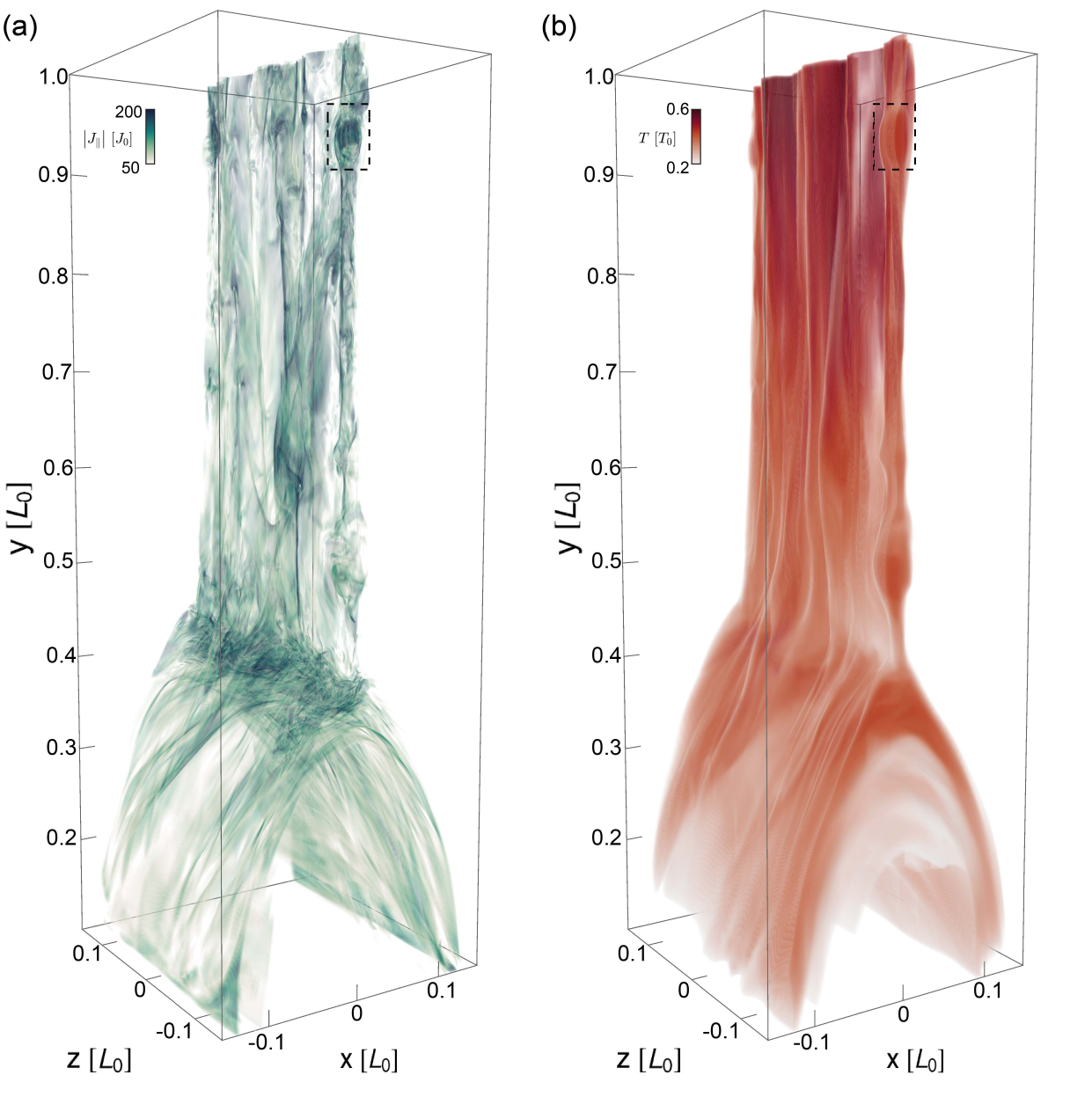}}
\caption{3D distributions of parallel current density (a), and temperature (b) for the current sheet (CS) and loop top at well-developed turbulent states at $t=8.2$ (reproduced from \citet{2025ren}).
The units of length, time, current density, and temperature are $L_0=50\,\mathrm{Mm}$, $t_0=114\,\mathrm{s}$, $J_0=9.54\,\mathrm{statC\,s^{-1}\,cm^{-2}}$, and $T_0=1.15\times 10^{7}\,\mathrm{K}$, respectively.}
\label{fig:mhd_overview}
\end{figure}

We use the high-resolution 3D MHD simulation of turbulent magnetic reconnection in the current sheet, originally reported by \citet{wangyulei2023}.
Based on the static mesh refinement (SMR) technique, the simulation adopts a high-resolution uniform mesh with a grid spacing of $\Delta L = 26\,\mathrm{km}$ to adequately capture the reconnection-driven turbulence within the CS and loop-top regions.
To compare, the entire simulation domain spans $50\,\mathrm{Mm} \times 100\,\mathrm{Mm} \times 15\,\mathrm{Mm}$ in the $x$, $y$, and $z$ directions, respectively, which corresponds to an effective grid number of $1920 \times 3840 \times 576$.
The simulation employs the HLLD Riemann solver \citep{2005Miyoshi}, which provides accurate resolution of MHD discontinuities with relatively low numerical diffusivity.
The most prominent feature of the MHD background employed is that strong turbulence is self-consistently developed across both the flare CS and loop-top region.
After the initiation of fast reconnection, the CS is quickly shattered by several types of instabilities, including the tearing mode instability (TMI), the kink instability, and the Kelvin–Helmholtz instability (KHI).
Finally, both the CS and loop top evolve into well-developed turbulent states.
Within the turbulent CS, thousands of fragmented reconnection patches are formed (see the parallel current density in Fig.\,\ref{fig:mhd_overview}a, also see \citet{2025ApJwyl}), which significantly heat and widen the CS structure (Fig.\,\ref{fig:mhd_overview}b).
By incorporating key thermodynamic processes—namely, thermal conduction, radiative cooling, and background heating—the synthetic images derived from our simulation data reproduce various observational signatures, including the broadening and fine-scale structures of the CS, as well as non-thermal broadening of spectral lines (see \citet{wangyulei2023} and \citet{2025ren}).
The development of turbulence in both the CS and loop-top regions has also been reported recently by \cite{2023ruan,2024ruan}, in which the KHI and Rayleigh–Taylor instability (RTI) are believed to be the key causes of turbulence.
The complex turbulent flows provide potential conditions for the formation of compression structures that might effectively accelerate particles.

\subsection{Modeling particle acceleration by Parker Transport Equation}
The acceleration and transport of electrons are investigated by solving PTE \citep{parker1965}:
\begin{equation}
    \frac{\partial f}{\partial t}+(\mathrm{\boldsymbol{V}}+\mathrm{\boldsymbol{V}_d})\cdot\nabla f-\frac{1}{3}\nabla\cdot\mathrm{\boldsymbol{V}}\frac{\partial f}{\partial\ln p}=\nabla\cdot(\boldsymbol{\kappa}\nabla f)+Q\,.
\label{eq:pte}
\end{equation}
Here $f(\mathbf{x},p,t)$ is the electron distribution function of space $\mathbf{x}$, momentum $p$, and time $t$, $Q$ denotes the particle source, 
$\mathrm{\boldsymbol{V}}$ is the plasma velocity of the MHD background,  $\mathrm{\boldsymbol{V}_d}=\frac{pcw}{3q} \nabla \times \left( \frac{\mathbf{B}}{B^2} \right)$ is the particle drift velocity, which incorporates both gradient and curvature drift.
It should be noted that PTE assumes an isotropic pitch-angle distribution, which is reasonable when particles are sufficiently scattered by strong turbulence \citep{2011daughton}.
The spatial diffusion tensor $\boldsymbol{\kappa}$ controls the transport of particles parallel and perpendicular to magnetic field lines.
To be specific, it is specified as $\kappa_{ij}=\kappa_{\perp}\delta_{ij}-(\kappa_{\perp}-\kappa_{\parallel})b_{i}b_{j}$, where $\kappa_{\parallel}$ and $\kappa_{\perp}$ are, respectively, the diffusion coefficients parallel and perpendicular to the local magnetic field direction $\mathrm{\boldsymbol{b}}$.
Here, we follow the method of \citet{jokipii1971,1999Giacalone} and evaluate the parallel diffusion coefficient by $\kappa_{\parallel}\approx1.622v^{4/3}L_c^{2/3}/(\Omega_0^{1/3}\sigma^2)$, where $v$ is the particle speed, $\Omega_0$ is the particle gyro-frequency, $L_c$ is the correlation length of turbulence, and $\sigma^2=\left\langle\delta B^2\right\rangle/B_0^2$ is the normalized wave variance of turbulence.
The turbulence correlation length $L_c$ is set as $500\,\mathrm{km}$, which is approximately the starting scale of the inertial region of the turbulent magnetic energy spectrum (see Fig.\,3b of \citet{wangyulei2023}).
$\sigma^2$ is thus set as 1, following the fully turbulent state in the MHD simulation \citep{wangyulei2023}.
The perpendicular diffusion coefficient is set as $\kappa_{\perp}/\kappa_{\parallel}=0.01$, guided by previous simulation results \citep{1999Giacalone}.

The transport equation is numerically solved using the stochastic differential equation (SDE) method, which maps the PTE into an equivalent set of SDEs and solves them by evolving pseudo-particles whose statistics represent the distribution function \citep{2010guo}.
The MHD background fields are saved at a fixed cadence of $\Delta t = 0.01\,t_0$, whereas the time step of the SDE pusher is constrained to remain smaller than the characteristic variation time scale of the MHD fields \citep[see also][and references therein]{2025li}.
For the time-evolving MHD background, the fields at intermediate particle times are approximated via linear interpolation between successive snapshots. 
The necessary MHD quantities at the particle positions are then obtained through trilinear interpolation using the field values at the eight surrounding grid cells.
The pseudo-particles are continuously injected at a constant rate into the heated regions in the CS, defined by $y\in\left[0.45,1\right]$ and $T\ge 3\,\mathrm{MK}$.
Their energies follow a Maxwellian distribution with a temperature of $\sim23\,\mathrm{MK}$ ($2\,\mathrm{keV}$) to mimic the high-temperature components in observations \citep{cx2018}.
The domain of the particle simulation is set as $x\in[-0.1,0.1]$, $y\in[0.3,1]$, and $z\in[-0.15,0.15]$, covering the regions of the CS and loop top.
The $x$ and $y$ directions employ open boundary conditions, whereas periodic boundaries are applied along the $z$ direction to mimic a long-extended flare loop (i.e., the direction of the polarity inversion line).
The pseudo-particles are excluded once they escape from the $\pm x$ or $\pm y$ boundaries.

We examine three distinct MHD background configurations. 
Runs 1 and 2 employ temporally static MHD backgrounds at two different moments.
Specifically, Run 1 uses MHD data at $t = 8.2$, when turbulence has already fully developed \citep{wangyulei2023}, whereas Run 2 uses data at $t = 7.5$, an earlier stage when turbulence has not yet fully developed, and a long-extended termination shock remains at the loop top.
In Run 3, we adopt a time-dependent MHD background that evolves dynamically from $t = 6.5$ (refreshing it every $0.01t_0$), covering the transition from a quasi-2D laminar system to a full 3D turbulent one.
For all three cases, particle simulations are terminated after a duration of $t = 3t_0$ (5.7 minutes), when the acceleration saturates, as indicated by the electron energy spectrum reaching a quasi-steady state.
It should be noted that, for Run 3, after $t=8.2$ the simulation is continued for an additional $1.3t_0$ using the static MHD fields at $t = 8.2$, to complete a full simulation time of $3t_0$ for comparison with the other two cases.

\begin{figure}
\centering
\resizebox{\hsize}{!}{\includegraphics{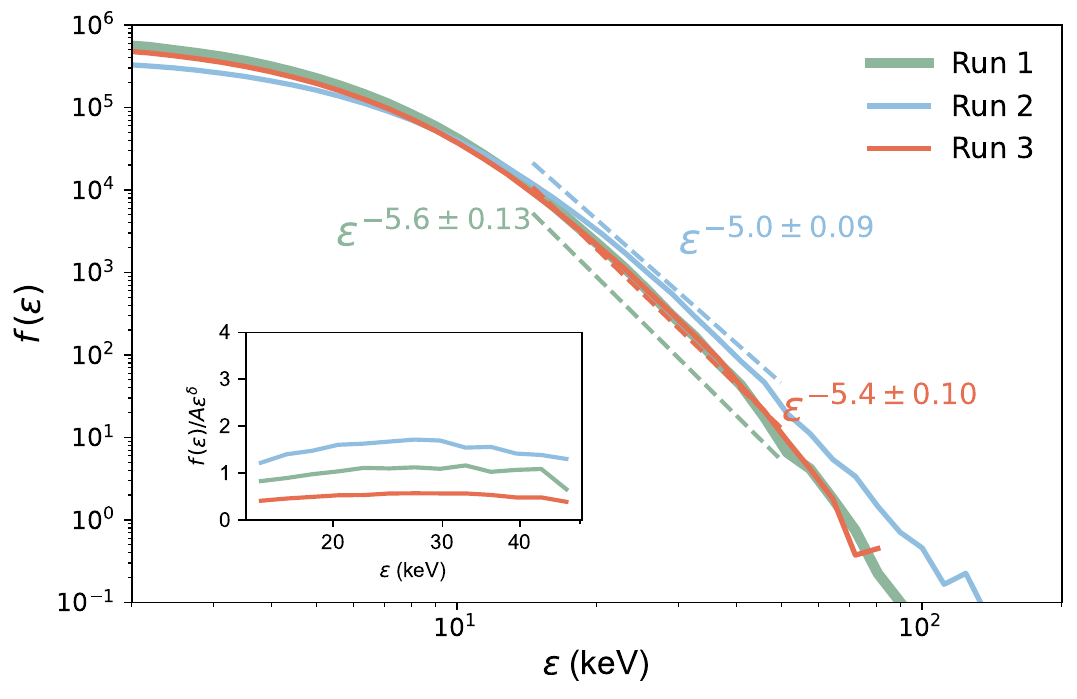}}
\caption{Energy spectra of electrons evaluated in the entire simulation domain for three cases.
The dashed lines show the power-law fits over the energy range of $15\,\mathrm{keV}$--$55\,\mathrm{keV}$.
The inset depicts the compensated plots $f(\epsilon)/\mathrm{A}\epsilon^{\delta}$ for three runs, where $\delta$ denotes the fitted power-law index.
}
\label{fig:three_contrast}
\end{figure}

\section{Results}\label{sec:result}
\subsection{Overview of electron acceleration}
For all three runs, initial thermal electrons can be accelerated to energies of up to $\sim90\,\mathrm{keV}$, with the resulting spectra exhibiting clear power-law features (see Fig.\,\ref{fig:three_contrast}).
It should be noted that the energy windows used for fitting the power-law function are comparable with those used for fitting the non-thermal components in solar HXR observations \citep{2004Grigis,2015oka,2017Aschwanden}.
Within these energy windows, the compensated curves are approximately horizontal, indicating a reliable power-law fit.

In Run 2, the system remains quasi-2D, featuring a long-extended termination shock at the loop top that provides an efficient acceleration site for electrons, yielding a power-law spectral index of $\sim-5.0$ over the energy range of $15\,\mathrm{keV}$--$55\,\mathrm{keV}$.
In contrast, Run 1 enters a turbulent state in both the CS and the loop-top regions, characterized by highly fragmented compression structures, corresponding to a spectral index of $\sim-5.6$.
Run 3 features an evolving MHD background, which incorporates the effects of both the early quasi-2D configuration and the later turbulent regime, giving rise to an intermediate power-law index.
Overall, the acceleration results are very similar across all three cases with comparable power-law spectral indices and maximum electron energies.
Hereafter, we focus on the results of Run 1 to explore the mechanism and capability of electron acceleration under a fully-developed turbulent state.

\begin{figure*}
\centering
\resizebox{\hsize}{!}{\includegraphics{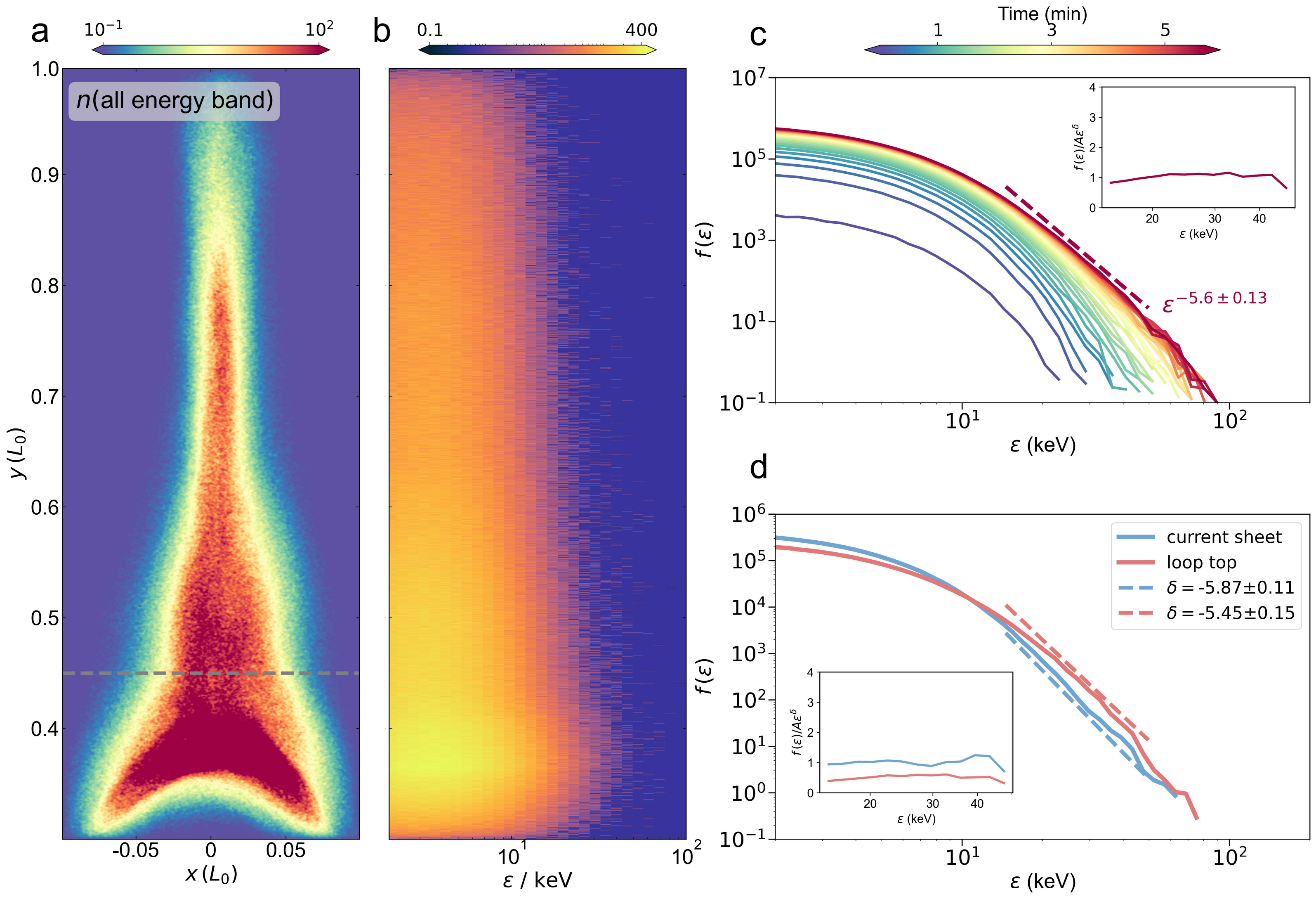}}
\caption{Spatial distribution and energy spectra of accelerated electrons from Run 1.
a: electron distribution in the $x$-$y$ plane, integrated along the $z$-direction.
The gray dashed line marks the rough boundary between the CS and the loop-top region.
b: the energy spectrum of electrons along the $y$–direction integrated over $x\in\left[-0.1,0.1\right]$.
c: energy spectra of all electrons in the entire simulation domain at different moments.
d: energy spectra at the CS (blue curve) and loop-top (pink curve) regions at the final moment. 
The dashed lines denote the power-law fitting, and the fit range is the same as in Fig.\,\ref{fig:three_contrast}.
The insets in c and d depict the corresponding compensated plots $f(\epsilon)/\mathrm{A}\epsilon^{\delta}$.
}
\label{fig:energy_spectra}
\end{figure*}

Figure \ref{fig:energy_spectra}a shows that the energetic electrons appear in both the flare CS and loop top at the end of the simulation, with the number density and the maximum energy peaking in the latter and decreasing with height (Fig.\,\ref{fig:energy_spectra}b).
The temporal evolution of the energy spectra of electrons clearly reveals the acceleration process, changing from a Maxwellian at the beginning to a non-thermal distribution after $3t_0$ (Fig.\,\ref{fig:energy_spectra}c).

We further separate the simulation domain into the loop-top ($y\in[0.35,0.45]$) and the CS ($y\in[0.45,1]$) regions to compare the acceleration results in the different regions where magnetic structures and dynamics are distinct.
Nevertheless, the interesting result is that the energy spectra for the non-thermal tails in both regions exhibit similar power-law indices ($-5.87$ in the CS, $-5.45$ at the loop top), although the population of energetic electrons above $15\,\mathrm{keV}$ in the CS is fewer--by about 45\%--than that at the loop top (see Fig.\,\ref{fig:energy_spectra}d).
This result indicates that the CS itself is capable of directly accelerating electrons, thus providing the possibility of locally producing HXR and microwave emissions, as observed previously \citep{2018gary,2020NAchenbin,2022kou}.

\subsection{Mechanisms responsible for acceleration}
\begin{figure*}
\centering
\resizebox{\hsize}{!}{\includegraphics{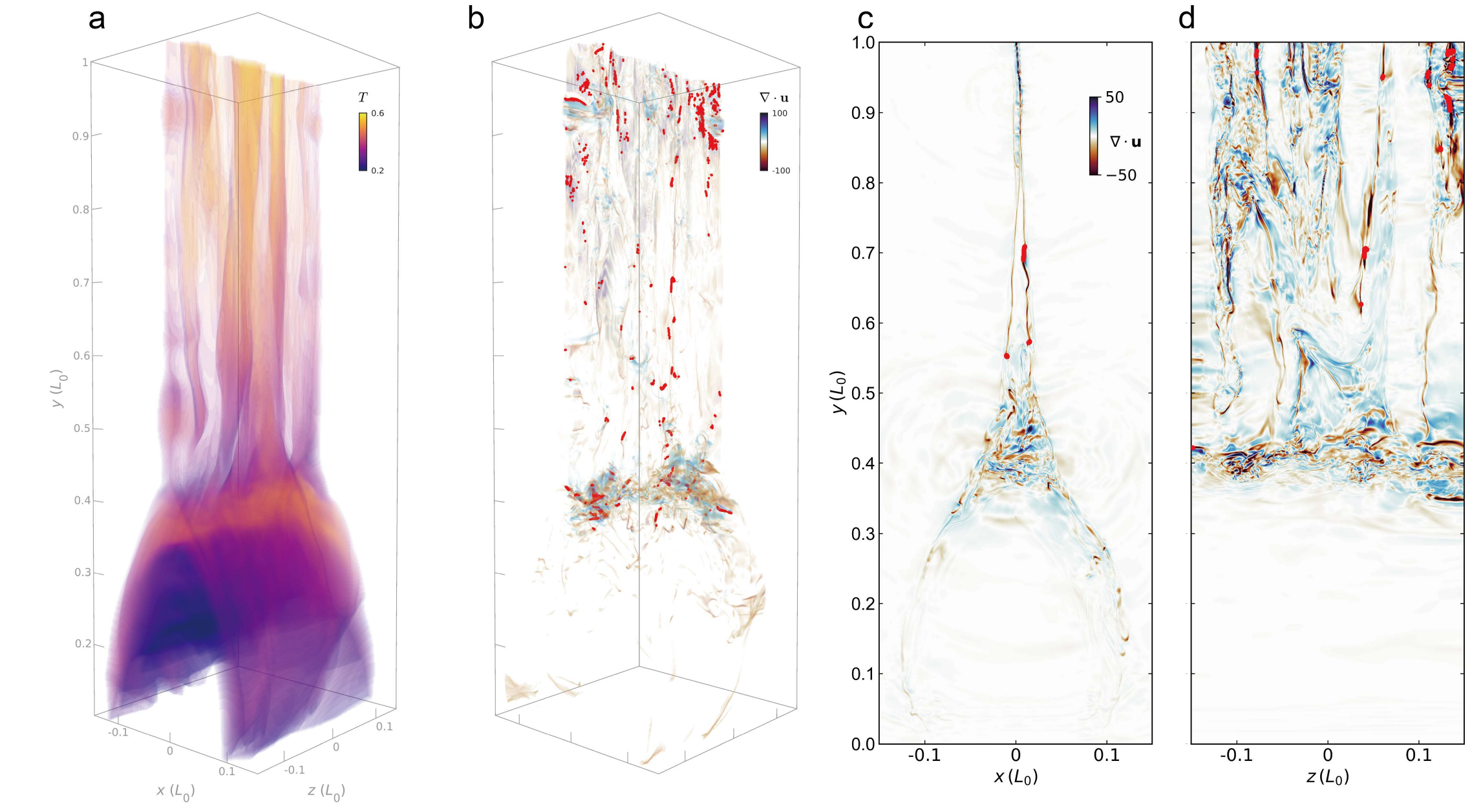}}
\caption{a and b: 3D distributions of temperature $T$ and velocity divergence $\nabla\cdot\mathbf{u}$.
c and d: two-dimensional (2D) distributions of $\nabla\cdot\mathbf{u}$ in the $x$-$y$ and $z$-$y$ plane, respectively.
The red dots in panels b--d indicate the locations of fragmented shocks.}
\label{fig:mhd3D2Dslice}
\end{figure*}

The electron acceleration is related to the fragmented compression structures driven by turbulent reconnection that appear within both the CS and the loop top. 
As shown by Fig.\,\ref{fig:mhd3D2Dslice}a, the CS and loop top are filled with high-temperature plasmas resulting from reconnection. 
The reconnection also forms outflows of different velocities and directions, which collide with each other to generate numerous compression structures as denoted by the divergence of velocity $\nabla\cdot\mathbf{u}$ in Fig.\,\ref{fig:mhd3D2Dslice}b.
Some regions with strong compression even form fragmented shocks (see the red dots marked in Fig.\,\ref{fig:mhd3D2Dslice}b, c, and d), distributed not only at the loop top but also throughout the CS.
In Appendix \ref{appendix:A}, we introduce the method for shock identification and provide an example of a fragmented shock with a compression ratio $\sim 2$ found in the CS region.

In contrast to previous 2D or 2.5D results, the CS in our 3D simulation is significantly broadened by magnetic turbulence \citep{2025ren}. 
As a result, the compression structures within it exhibit a highly disordered spatial distribution (Fig.\,\ref{fig:mhd3D2Dslice}c).
Obviously, this significantly differs from the classical Petschek-type configuration characterized by a single X-point and two pairs of slow-mode shocks \citep{1964Petschek}.
The acceleration mechanism at the flare loop top is also found to be markedly different from the termination shock scenario. 
As shown by the slice of negative $\nabla\cdot\mathbf{u}$ in the $z$-$y$ plane (Fig.\,\ref{fig:mhd3D2Dslice}d), no counterpart of the 2D loop-top termination shock typically appearing in standard-flare acceleration models is found along the $z$-direction.
Instead, the turbulence destroys the expected large-scale termination shock and gives rise to numerous localized and strongly fragmented compression structures, even containing fragmented shocks, that dominate the acceleration of electrons.

\subsection{Trajectories of energetic electrons}
\begin{figure*}
\centering
\resizebox{\hsize}{!}{\includegraphics{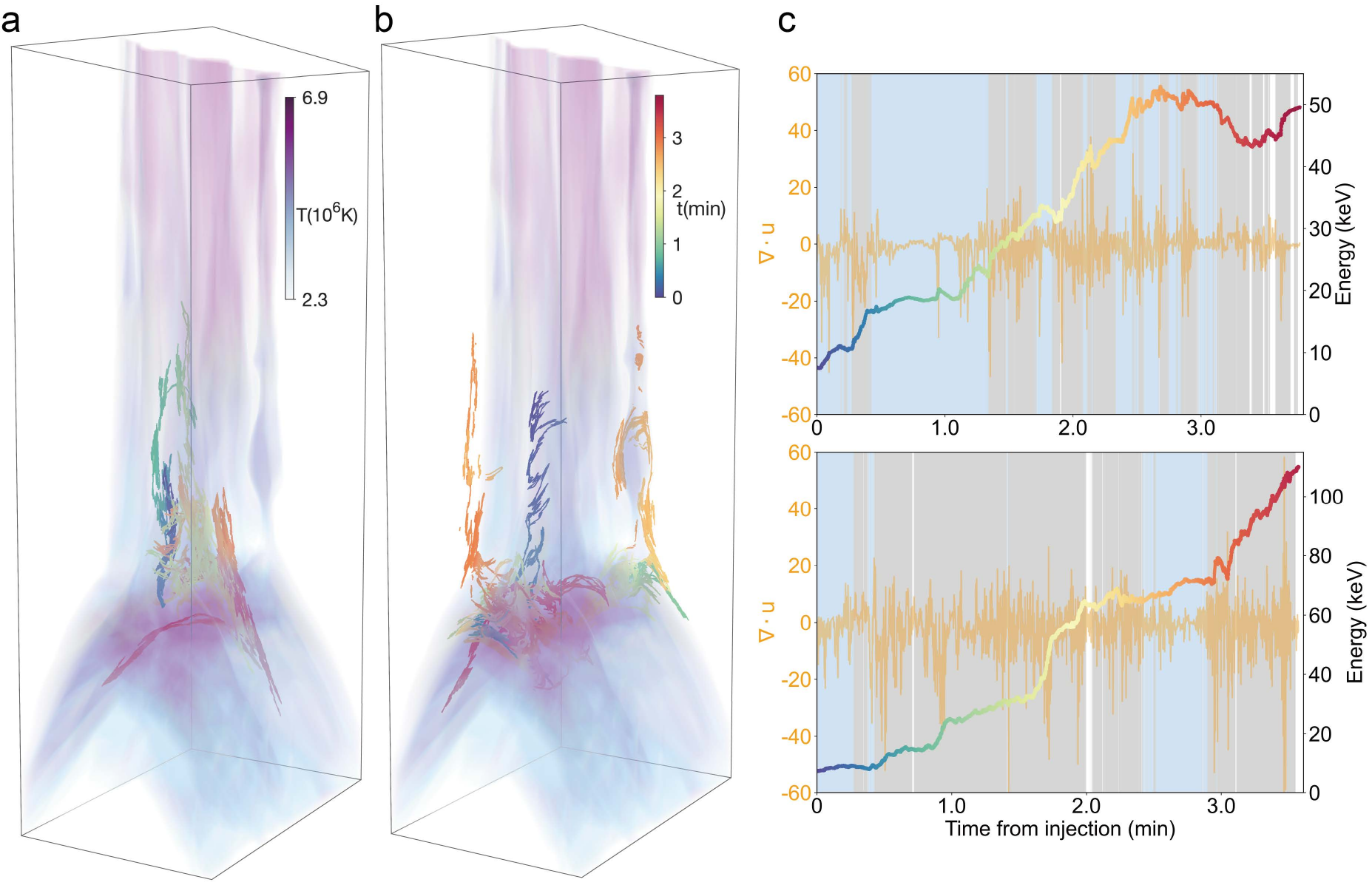}}
\caption{a and b: 3D trajectories of two representative pseudo-particles in Run 1, overplotted on the temperature distribution and color-coded by time.
c and d: Temporal evolution of particle energy along the corresponding trajectories, overlaid with the local compression strength along their paths (averaged every 20 particle steps).
The background shading indicates particles located in the CS (blue) or the loop-top region (gray).
}
\label{fig:pic_trajectory}
\end{figure*}

Tracing the trajectories of accelerated electrons can provide valuable information about the acceleration processes in a 3D turbulent environment.
Fig.\,\ref{fig:pic_trajectory}a and Fig.\,\ref{fig:pic_trajectory}b show two representative electron trajectories in Run 1: one shows comparable acceleration in both the loop top and CS, and the other is dominated by the acceleration in the loop-top region. 
The paths display complex 3D motions, including strong spreading motions along the $z$-direction rather than simple up-and-down motion, emphasizing substantial 3D effects on particle transport.
In particular, a complex 3D flux rope structure within the CS imposes strong confinement on electron motion (see the orange feature in Fig.\,\ref{fig:pic_trajectory}b), emphasizing the critical role of 3D magnetic reconnection in particle transport.
Within the loop-top region, the electron trajectories remain highly complex due to magnetic turbulence driven by intermittent and fragmented reconnection outflows, thus enhancing the confinement of electrons relative to that in the CS (as shown in Fig.\,\ref{fig:pic_trajectory}b).

Fig.\,\ref{fig:pic_trajectory}c illustrates the temporal evolution of the energy of representative electrons, overlaid with the variation of local $\nabla\cdot \mathbf{u}$.
One can find that the acceleration of electrons is due to their crossing fragmented compression structures (some sharp peaks in the $\nabla\cdot \mathbf{u}$ curve) numerous times; the energy gained each time is inferred from the compression term of the PTE.
Moreover, the energy variation of electrons along the trajectories shows that each electron is accelerated in both the CS and loop-top regions, with their relative contributions varying from case to case. 
Some electrons even attain higher energy in the CS than in the loop top, as shown by the first trajectory.
Furthermore, although electrons sometimes lose energy when crossing expansion regions, the repetitive crossings of electrons within both regions result in a considerable total energy gain. 
Therefore, it is suggested that turbulent magnetic reconnection can efficiently accelerate electrons to high energy through turbulence-driven fragmented compression structures during solar flares.

\subsection{Acceleration capabilities of CS and loop top}
To investigate the acceleration capabilities of the CS and loop top, we track the trajectories of a sample of 1500 electrons finally exceeding $30\,\mathrm{keV}$.
Among these samples, only 25\% are accelerated exclusively at the loop top.
The remaining majority (75\%) gain energy in both the CS and loop-top regions.
For our subsequent analysis, we focus on this latter subset.
Within it, the total energy acquired in the CS constitutes approximately 21\% of that gained at the loop top, quantitatively highlighting the difference in acceleration efficiency between the two regions. 
To further characterize the acceleration process of individual electrons, we record the initial energy ($E_0$), the final energy ($E_{\mathrm{final}}$), and the total energy increments $\delta E_{cs}$ and $\delta E_{lt}$ as obtained in the CS and loop top, respectively.
It should be noted that all sampled electrons have initial energies exceeding $1.9\,\mathrm{keV}$, and $90\%$ of their $E_0$ values are above $5\,\mathrm{keV}$. 
This is consistent with the assumption of DSA theory, which requires that particles attain a sufficiently high injection energy before being accelerated.

\begin{figure*}
\centering
\resizebox{\hsize}{!}{\includegraphics{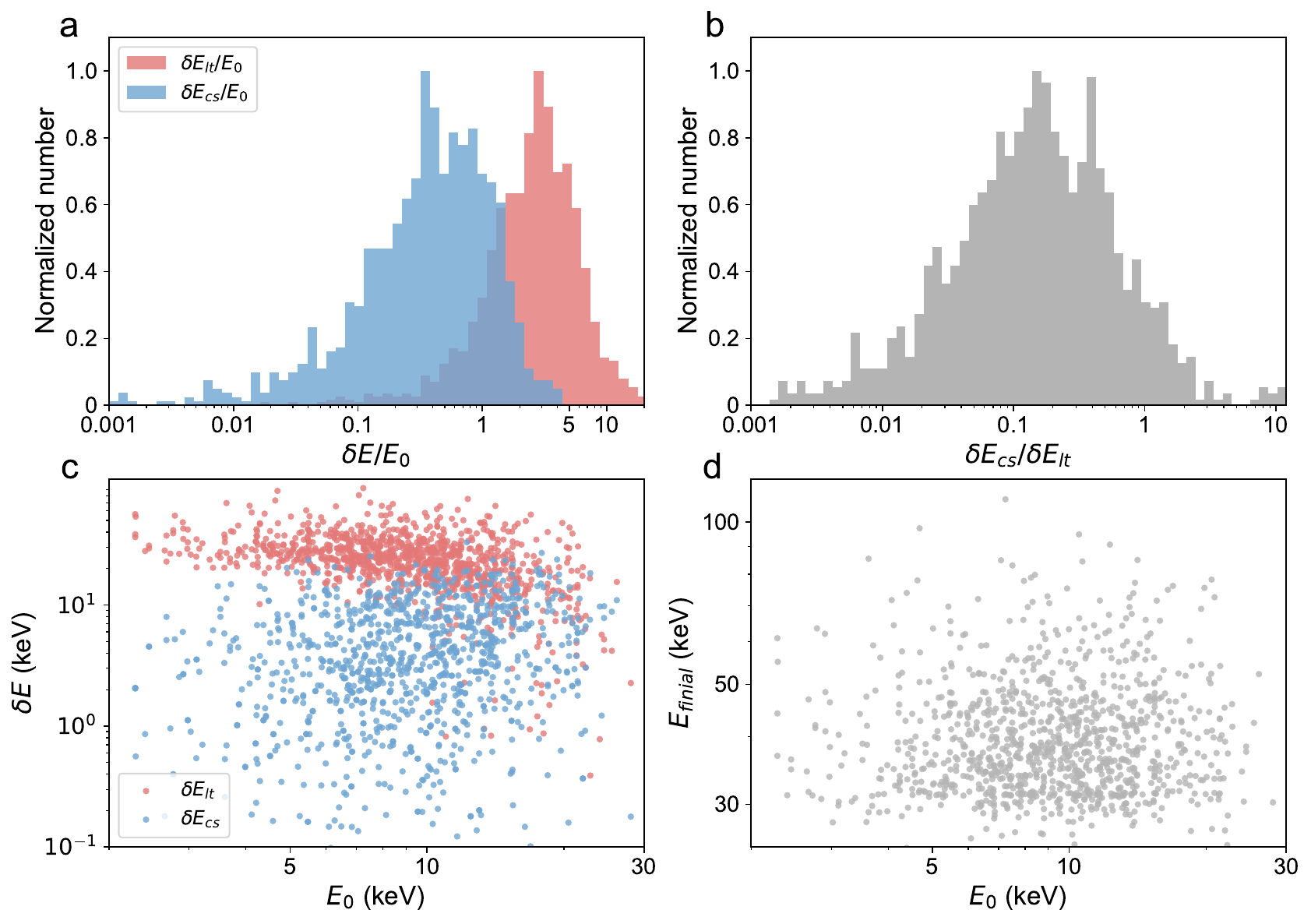}}
\caption{
a and b: histograms of acceleration ratios $\delta E/E_0$ in both regions and the ratio of energy gains in the CS to that in loop-top regions ($\delta E_{\mathrm{cs}}/\delta E_{\mathrm{lt}}$).
c and d: Scattering plots of the energy increment $\delta E$ and the final energy $E_{final}$ vs. the initial energy $E_0$, respectively.
}
\label{fig:statistics}
\end{figure*}

Fig.\,\ref{fig:statistics}a shows the distribution of the ratio of the energy increment in both regions to the initial energy, $\delta E_{cs,lt}/E_0$. 
One can find that the electrons typically gain 30\%–40\% of their initial energy in the CS, while they gain energy of several times $E_0$ at the loop top and even exceed an order of magnitude in some cases.
This demonstrates that the electrons are accelerated most efficiently at the loop top.
Fig.\,\ref{fig:statistics}b further displays the distribution of the ratio of the energy increment in the CS to that in the loop top, $\delta E_{cs}/\delta E_{lt}$. 
Although it peaks between 10\% and 20\%, it also extends to 100\% and higher values, indicating that a minority of electrons can attain more energy in the CS. 
This is consistent with the two acceleration patterns identified by the example trajectories: comparable LT–CS acceleration for a smaller fraction and loop-top–dominated acceleration for most electrons (see Fig.\,\ref{fig:pic_trajectory}a and Fig.\,\ref{fig:pic_trajectory}b).

We also examined the influences of the initial energy ($E_0$) on the energy increments ($\delta E_{cs,lt}$) and the finally-reached energy ($E_{final}$) in Figs.\,\ref{fig:statistics}c and \ref{fig:statistics}d, respectively.
In the CS, the energy increment exhibits a highly scattered distribution with $E_0$ on a logarithmic scale, while in the loop top, it becomes noticeably more compact; though it still approaches a nearly flat distribution.
Both display that the energy increments are almost independent of the initial energy in the 3D turbulent acceleration scenario.
Moreover, for the same $E_0$, electrons generally display larger energy increments in the loop top than in the CS.
These results are expected because, for a given $E_0$, electrons in the loop top generally cross stronger fragmented shocks and have longer-term confinement, thus achieving higher acceleration ratios, particularly for low-energy electrons.

Finally, it is found that the final energies are also independent of initial energies (Fig.\,\ref{fig:statistics}d). 
That is to say, electrons with similar initial energies can obtain various energies, and electrons with similar final energies may originate from distinct initial energies.
It further supports that all electrons undergo totally random acceleration and transport histories in the 3D turbulent environment.

\section{Conclusions and Discussion}\label{sec:conclusions and discussion}
In this work, we investigate electron acceleration during the turbulent reconnection of solar flares.
We demonstrate that the primary acceleration is achieved by the turbulence-driven compression structures, including fragmented shocks, pervading both the flare CS and loop top.
Even in the absence of a coherent large-scale termination shock, electrons can still be efficiently accelerated at the loop top.
More importantly, we find that the CS has considerable acceleration capability compared to that of the loop top, with the total energy gained by all electrons from the CS reaching 21\% of that from the loop top, and some electrons even acquiring more energy in the CS than at the loop top.
To be specific, both regions can efficiently accelerate electrons up to $\sim90\,\mathrm{keV}$, and the corresponding energy spectra present a power-law distribution with an index of $\sim-5.5$, similar to some observation results \citep{2015oka}.
It is also found that the initial energy of accelerated electrons has weak relevance to both the energy increment and the final energy, strongly indicating a random, unpredictable process of acceleration and transport within the strongly turbulent reconnection.

Electron acceleration in solar flares has been argued to be primarily due to the loop-top termination shock, with non-thermal emissions in the CS interpreted as a secondary transport effect \citep{2024chen}.
However, such a termination shock predicted by 2D models is likely to be disrupted in 3D turbulent systems.
Recently, \citet{2025li} explored particle acceleration in a 3D solar flare and reported electron acceleration within the CS, wherein the CS remains quasi-2D, and the loop-top termination shock still exists, likely due to the insufficient development of turbulence.
In our model, the well-developed strong turbulence in the CS is capable of forming large amounts of fragmented compression structures and even fragmented shocks, which thus provide a new way to locally accelerate thermal electrons and generate non-thermal emissions at the CS.
Compared with conventional stochastic acceleration scenarios \citep{1997miller,2012Petrosian}, fragmented compression structures in our model play a dominant role in particle acceleration, indicating a qualitatively different acceleration mechanism operating during turbulent magnetic reconnection.
Our quantitative comparison of the energy gains in the CS and the loop-top regions further verifies that the role of the CS is not merely a passive transport channel but an efficient accelerator during solar eruptions.

Our results show that the initial electron energy has a limited effect on both the total energy gain and the final energy, implying a non-selective acceleration history in a highly turbulent system.
For the acceleration process at a single, stationary shock with a constant compression ratio and spatial scale, according to the PTE (Eq.\,\ref{eq:pte}), the energy gain rate $\mathrm{d}E/\mathrm{d}t$ is proportional to $E$.
Considering that the parallel diffusion coefficient $\kappa_\parallel$ scales as $\sim E^{2/3}$, we then have $\delta E \sim E^{1/3}$.
However, in the CS and the loop top of our simulation, numerous compression structures and shocks with varying properties exist, and the trajectories of particles across these acceleration elements can be highly random.
The collective effects of all the different compression structures passed by an electron can weaken the dependence of the total energy increments on the initial energy.

Beyond the electron acceleration, our model also offers a physically grounded explanation for the electron beam prescriptions that are commonly used to investigate the chromospheric spectra of flares.
It is found that the electrons that escaped from the bottom boundary also exhibit a power-law energy spectrum with a spectral index of $\sim-5.1$, extending to approximately $100\,\mathrm{keV}$ (see Fig.\,\ref{fig:escaped} in Appendix \ref{appendix:B}). 
The low-energy cutoff and spectral index are consistent with those of nonthermal electron populations assumed in 1D beam-driven simulations \citep{2015allred,2023Carlsson}.
Importantly, the resulting spectrum emerges directly from the simulated processes of electron acceleration and transport, which implies that those parameterised electron-beam spectra could physically arise from turbulent acceleration during flares.

Finally, several limitations of the current study are worth mentioning.
First, the application of the PTE implicitly assumes an isotropic distribution of electron momenta.
This assumption is expected to be reasonable because strongly turbulent structures can confine particles for a duration that permits isotropization. 
Future work will be conducted to quantitatively compare with more sophisticated models that account for pitch-angle anisotropy, such as the focused transport equation \citep{2014zank,2016leRoux}.
It should be noted that the PTE can account for the magnetic bottle effects even without explicitly modeling pitch-angle evolution, since the drift motions arising from magnetic gradient and curvature can change particle transport. \citet{2019kong} have shown that the magnetic-bottle configuration at the flare loop can trap particles and thereby enhance their acceleration.
Meanwhile, the PTE primarily focuses on diffusive shock acceleration (first-order Fermi).
A detailed study and quantitative comparison with the stochastic acceleration mechanism (second-order Fermi) proposed by \citet{2024Bacchini} and \citet{2026Mora} would require further dedicated work.
Second, we adopt diffusion coefficients comparable to those used in previous PTE–based flare studies (e.g., \citet{2019kong,2025li}). However, the expression of the coefficient is derived from quasi-linear theory \citep{jokipii1971,1999Giacalone}, and its validity in the regime of strong turbulence remains an open question that warrants further investigation.
Third, limited by computational resources, our MHD simulation box only covers part of the reconnection region along the polarity inversion line.
For a more comprehensive understanding of the electron acceleration processes, a new MHD simulation including high-resolution reconnecting CS and the entire CME–flare structure is urgently needed.
Finally, our work cannot quantify the absolute energy budget of nonthermal particles, because PTE framework assumes the nonthermal particles have little effects on the background and the pseudo-particles in solving the transport equation only reflect the relative distribution of particles.
Despite these limitations, our results provide a new pathway for understanding the acceleration of particles during solar flares and other energetic phenomena in the universe.

\begin{acknowledgments}
This research is supported by the National Natural Science Foundation of China under grants 12525305 and 12473057, the Fundamental Research Funds for the Central Universities (KG202506). 
The particle simulation is performed in the cluster system of the High Performance Computing Center (HPCC) of Nanjing University.
We would like to thank the anonymous referee for valuable suggestions.
\end{acknowledgments}

\bibliography{sample701}{}
\bibliographystyle{aasjournalv7}



\appendix
\restartappendixnumbering

\section{Shock identification}\label{appendix:A}
\begin{figure}
\centering
\includegraphics[width=12cm]{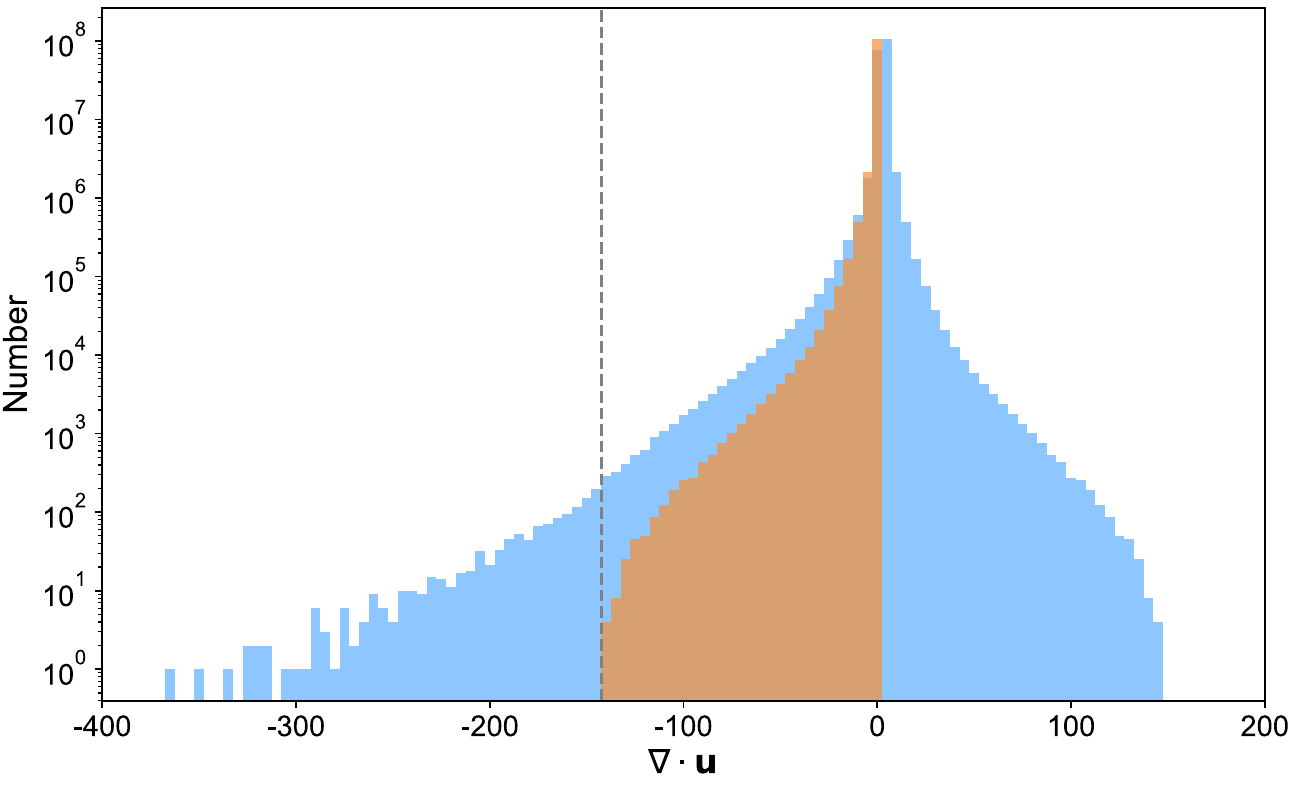}
\caption{
Histogram of $\nabla \cdot \mathbf{u}$ at grids in the CS region (blue shade).
The orange shade depicts the mirrored image of the $\nabla \cdot \mathbf{u} > 0$ part with respect to the $\nabla \cdot \mathbf{u}=0$ axis.
The vertical dashed line marks the location of $\nabla \cdot \mathbf{u}=-140$, the lower limit of the orange shade.
}
\label{fig:divv_hist}
\end{figure}

To identify shocks as shown in Fig.\,\ref{fig:mhd3D2Dslice}b--d, we first use the velocity divergence to select candidate locations following \citet{2020wangyikang}. 
Specifically, we construct the histogram of $\nabla\cdot\mathbf{u}$ within the CS region (see the blue-shaded area in Fig.\,\ref{fig:divv_hist}) and mirror its positive part to the negative side (see the orange shaded area in Fig.\,\ref{fig:divv_hist}). 
Since the symmetric part of the distribution about $\nabla\cdot\mathbf{u}=0$ is associated with linear wave components, grid points with $\nabla\cdot\mathbf{u}$ below the lower limit of the mirrored distribution ($\nabla\cdot \mathbf{u}<-140$) are selected as candidate shock locations.
The candidates are then verified using the Rankine--Hugoniot jump conditions.
Briefly, the local density gradient is used to determine the shock normal, and the upstream and downstream plasma parameters are obtained by interpolating the MHD variables along the normal direction. 
Only candidates satisfying the corresponding upstream and downstream conditions are retained as true shocks.
A similar method has also been used to identify shocks in 2D MHD simulations (e.g., \citet{2021snow}).

\begin{figure*}
\centering
\resizebox{\hsize}{!}{\includegraphics{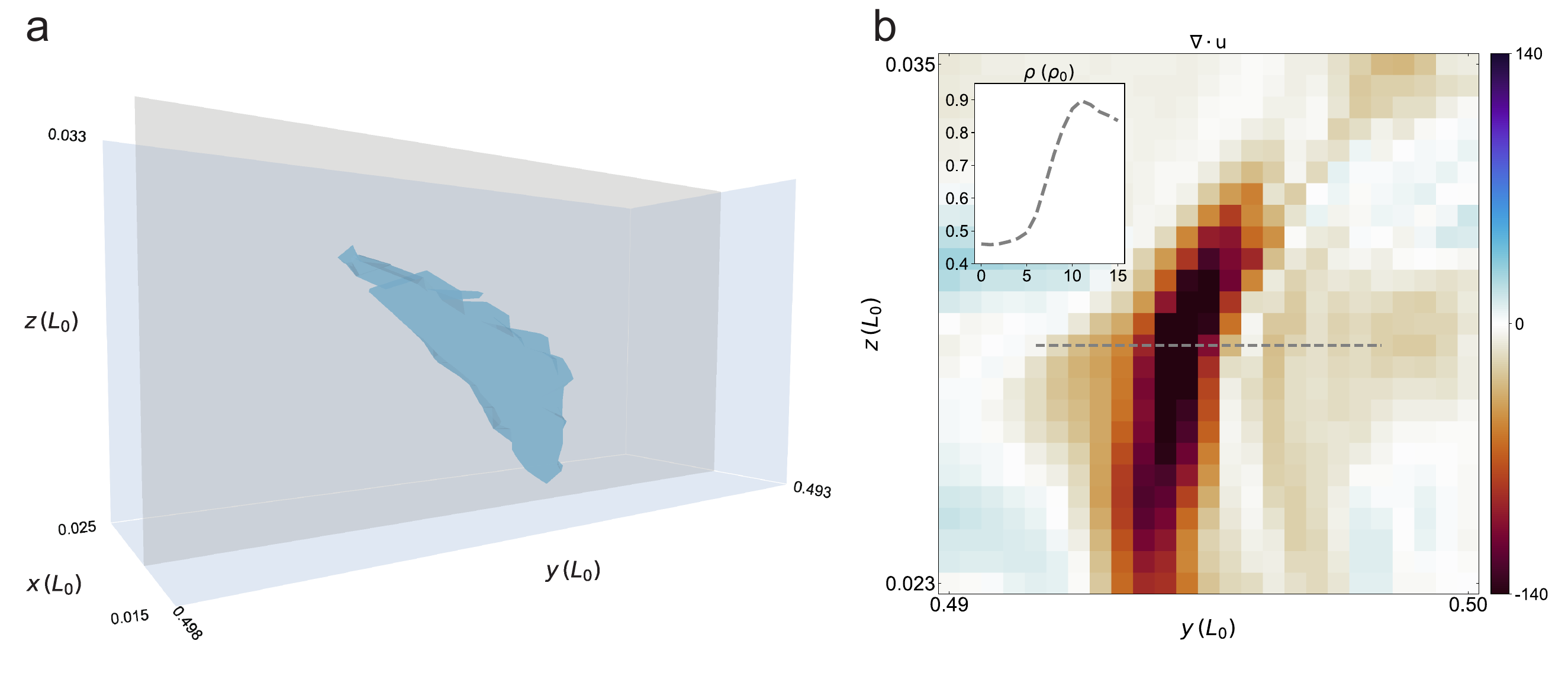}}
\caption{ A typical shock structure within the current sheet region. 
a: 3D distribution of the shock as shown by the blue isosurface of $\nabla \cdot \mathbf{u} = -140$.
The gray plane indicates the slice used by panel b.
b: Distribution of $\nabla \cdot \mathbf{u}$ on the slice across the shock. 
The grey dashed slit traverses the upstream and downstream regions; the inset shows the profile of mass density along this slit.}
\label{fig:shock}
\end{figure*}

Figure \ref{fig:shock} presents an example of a shock identified within the CS region. 
The shock is visualized by the isosurface of $\nabla\cdot\mathbf{u} = -140$ (Fig.\,\ref{fig:shock}a), which consists of multiple grid cells and therefore indicates that shocks are spatially resolved in the turbulent conditions.
The mean density-gradient direction of the shock patch is $(0.15,0.88,-0.36)$, giving the approximate shock normal, which is nearly aligned with the $y$-axis. 
The corresponding $\nabla\cdot\mathbf{u}$ distribution and density profile across the shock are shown in Fig.\,\ref{fig:shock}b, giving a density compression ratio of $\sim2$.

\section{Energy spectra of escaped electrons}\label{appendix:B}
We collect the electrons escaping through the lower boundary from $t=1.8t_0$ to $t=2.8t_0$ in Run 1, a relatively late stage when the acceleration and transport processes are well established, and construct their energy spectrum.
The resulting spectrum follows an approximate power-law distribution over the tens-of-keV energy range, with a spectral index of $\delta\sim-5.1$.

\begin{figure}
\centering
\includegraphics[width=12cm]{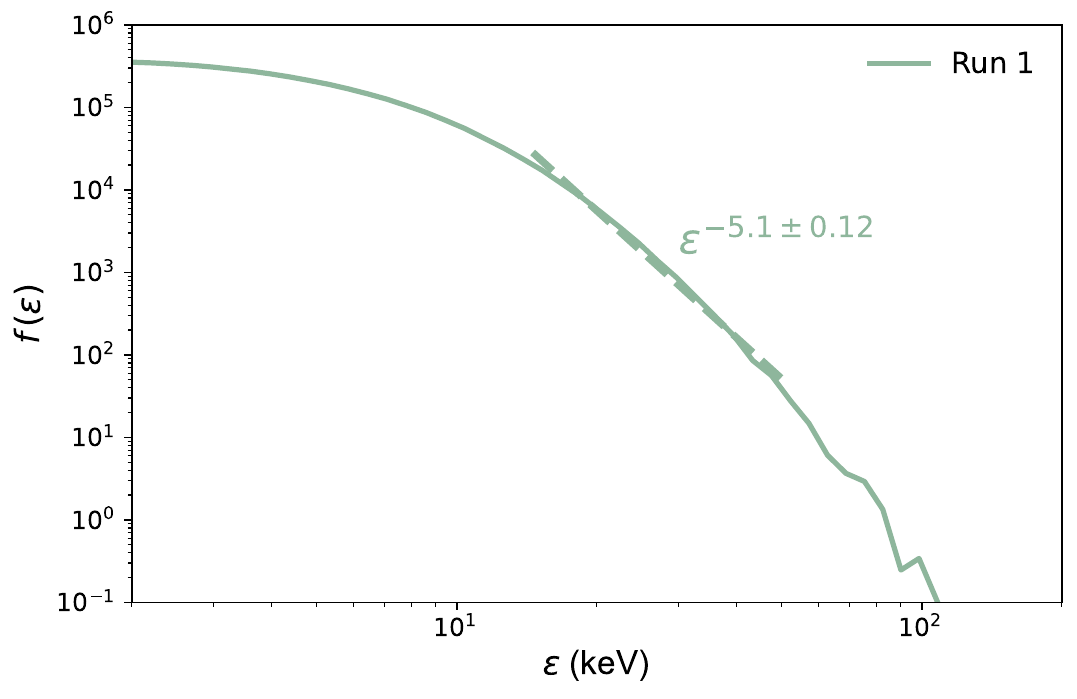}
\caption{
Energy spectrum of the electrons escaping through the lower boundary of the simulation domain during the interval from $t=1.8t_0$ to $t=2.8t_0$. 
The dashed line denotes the power-law fit, giving a spectral index $\sim-5.1$.
}
\label{fig:escaped}
\end{figure}

\end{document}